# Secure and Scalable Network Slicing with Plug-and-Play Support for Power Distribution System Communication Networks

Jian Zhong, Chen Chen, *Senior Member, IEEE*, Yuqi Qian, Yiheng Bian, Yuxiong Huang, and Zhaohong Bie, *Fellow, IEEE*

*Abstract*—With the rapid development of power distribution systems (PDSs), the number of terminal devices and the types of delivered services involved are constantly growing. These trends make the operations of PDSs highly dependent on the support of advanced communication networks, which face two related challenges. The first is to provide sufficient flexibility, resilience, and security to meet varying demands and ensure the proper operation of gradually diversifying network services. The second is to realize the automatic identification of terminal devices, thus reducing the network maintenance burden. To solve these problems, this paper presents a novel multiservice network integration and device authentication slice-based network slicing scheme. In this scheme, the integration of PDS communication networks enables network resource sharing, and recovery from communication interruption is achieved through network slicing in the integrated network. Authentication servers periodically poll terminal devices, adjusting network slice ranges based on authentication results, thereby facilitating dynamic network slicing. Additionally, secure plug-and-play support for PDS terminal devices and network protection are achieved through device identification and dynamic adjustment of network slices. On this basis, a network optimization and upgrading methodology for load balancing and robustness enhancement is further proposed. This approach is designed to improve the performance of PDS communication networks, adapting to ongoing PDS development and the evolution of PDS services. The simulation results show that the proposed schemes endow a PDS communication network with favorable resource utilization, fault recovery, terminal device plug-and-play support, load balancing, and improved network robustness.

## I. Introduction

WITH the wide application of information and communication technology (ICT), the modernization of power systems is continuing at a rapid pace. In particular, power distribution systems (PDSs),

This work was supported by the Science and Technology Project of State Grid Corporation of China under Grant 5400-202199524A-0-5-ZN. (Corresponding author: Chen Chen.)

Jian Zhong, Chen Chen, Yuqi Qian, Yiheng Bian, Yuxiong Huang, and Zhaohong Bie are with the State Key Laboratory of Electrical Insulation and Power Equipment, Shaanxi Key Laboratory of Smart Grid, Xi'an Jiaotong University, Xi'an 710049, China (e-mail: zhongjian0829@stu.xjtu.edu.cn; morningchen@xjtu.edu.cn;qianyuqi77@stu.xjtu.edu.cn;byh0xyz@stu.xjtu.edu.cn; yuxionghuang@xjtu.edu.cn; zhbie@mail.xjtu.edu.cn).

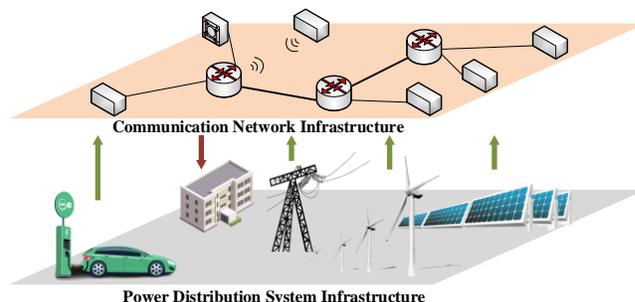

**Fig. 1.** The relationship between the PDS infrastructure and the ICT infrastructure.

as the parts of power systems that receive electricity from power transmission or generation infrastructure and distribute it locally through distribution facilities to various types of users, are evolving at the fastest rate [1]. The associated development is leading to the inclusion of an increasing number of field terminal devices participating in diversified services (such as advanced metering, renewable energy generation, and transportation electrification) in PDSs [2].

The above phenomenon poses challenges to PDS communication networks with respect to reliability, responsiveness [3], and scalability [4]. First, the wide integration of ICT into PDSs has led to a high dependency of PDSs on communication networks. As shown in Fig. 1, efficient PDS management is based on the ability to monitor and control PDS facilities in real time, which requires a large amount of information collected from terminal devices over communication networks [4]. Therefore, the growing number of devices and the diversification of services impose high scalability requirements on PDS communication networks. Moreover, the workload of interfacing with and managing terminal devices as well as the allocation of network resources to these terminal devices place high demands on device manageability in these communication networks. Second, PDS terminal devices tend to be geographically distributed [5] in public spaces (e.g., utility poles, houses, or buildings) [6], increasing the vulnerability of PDS communication networks to cyberattacks and unintentional errors [7]. Due to the reliance of PDSs on communication networks, communication network failures can severely disrupt normal PDS operations and may even trigger recursive cascading of outages across PDSs [8]. PDS utilities in Ukraine and Venezuela suffered cyberattacks in 2015 and 2019, respectively, resulting in massive power outages leading to multibillion-dollar losses [6], [9]. Addressing these problems requires PDS communication networks to be cybersecure and resilient. Therefore, a PDS communication network that is secure [10], scalable [4], fault tolerant [11], capable of dynamically interfacing with terminal devices and ensuring fast recovery in cases of failure is needed.

Network slicing technology within software-defined networking (SDN) architectures offers several benefits for addressing these issues [12], [13]. SDN technology decouples the control and forwarding functions of network devices, abstracting network services from the underlying hardware devices and allowing software-defined controllers to achieve network intelligence [14]. Network slicing is one of the most popular network virtualization technologies for allocating and isolating network resources [15]. This technology creates a way for various network services to share the same physical network without knowing its underlying details by abstracting and partitioning this physical network into multiple parts [16]. In a network slicing architecture, each of the logical network slices is configured to provide specific network capabilities and features [15]. Dynamic resource allocation between slices makes it possible to meet the quality of service (QoS) requirements of different applications under various scenarios. It has been shown that network resilience and security can be achieved through dynamic resource allocation and isolation by means of network slicing [17]. Therefore, the implementation of network slicing in PDS communication networks offers important benefits for meeting the requirements of reliable, scalable, secure, and resilient network functions.

Research on network slicing has contributed to improvements in network performance in terms of resource allocation, QoS improvement, traffic aggregation, etc. However, the following difficult problems arise when constructing network slices for PDS communication networks:

1) *Concentrated Maintenance:* PDSs are typically connected to many terminal devices distributed over a wide area via communication networks. Existing network maintenance practices often rely on manual on-site tuning, which can be labor intensive and inefficient. Thus, there is great interest in creating a system that supports dynamic terminal device access maintenance, thereby reducing the workload for personnel and improving network management efficiency.

2) *Dynamic Identification:* PDSs are responsible for providing power to consumers, so the security of PDS networks is of critical importance. The ability to efficiently identify devices and verify their network access authority after they interface into the network is essential for mitigating the network security risks posed by unauthorized access.

3) *Network Risk Isolation:* The increasing use of digital and decentralized technologies in PDSs provides a large cyberattack surface in communication networks [40]. To protect PDS communication networks from illegitimate information access and network service disruption attacks, such as denial of service (DoS) and false data injection (FDI) attacks, it is highly important to find a way to isolate terminal devices before they are successfully authenticated.

4) *Existing Equipment Utilization:* Existing PDS networks contain a variety of wired and wireless communication devices. For economic reasons, the use of existing resource-constrained legacy equipment should be considered when integrating new technologies. Accordingly, the network slicing scheme should be compatible with both wireless and wired transmission paths and should make full use of the resources of the existing network facilities.

5) *Network Architecture Optimization:* As the number of terminal devices participating in PDS communication networks continues to increase, the bandwidth utilization of the existing communication networks will gradually become saturated, and the network robustness will further decrease. Thus, there is a need for network evaluation and optimization methods based on service requirements. The formulation of an economical optimization scheme for load balancing and improving the robustness of existing networks is therefore a crucial issue for the development of PDS communication networks.

To address the above challenges, we propose a device authentication slice-based network slicing scheme. In particular, we integrate multiple different service networks to achieve resource sharing and communication failure resilience with minimal changes to the original communication networks. Moreover, the establishment of a device authentication network slice with its own authentication server in the integrated network is proposed. In this scheme, different service slices are established in accordance with diverse service requirements. Terminal devices are permitted to connect only to the authentication slice before they are successfully authenticated, and can access their corresponding network slices only after they are authenticated by the server. This enables terminal devices to interact with the PDS communication network in a secure plug-and-play manner and isolates PDS services from unauthorized access. Furthermore, the range of service slices can be dynamically adjusted in accordance with the device authentication results, providing security and adjustability for PDS network services. Finally, we build a PDS communication network evaluation model to evaluate the robustness of the network nodes and links. This model can be further used to construct a load-balanced and robustness-enhanced network optimization plan with the lowest cost based on the network structure and service requirements, thereby helping to optimize PDS communication networks to continually adapt to evolving service requirements. The main contributions of this paper are summarized as follows.

1) To our knowledge, this is the first paper to propose a device authentication slice-based network slicing scheme. This scheme can realize secure plug-and-play support for terminal devices while improving the security of a PDS communication network by isolating unauthorized devices in the authentication slicing layer and adjusting the range of network slices based on the device

2) To determine the network slicing scheme, we formulate the network slicing problem as a two-stage mixed-integer linear programming (MILP) model. This model is able to quickly provide network slicing solutions for PDS communication networks, thereby facilitating the dynamic adjustment of PDS network service slices.

3) We propose a PDS network optimization model aimed at load balancing and robustness enhancement. This model can be used to evaluate network robustness and derive the lowest-cost architectural optimization and updating solution that meets specified robustness and load balancing requirements.

The remainder of this article is organized as follows. Section II provides an overview of related work. Section III describes the proposed network slicing method and corresponding model. Section IV then formulates the models for evaluating and improving PDS network load balancing and robustness. Numerical case verifications are presented in Section V, and conclusions are drawn in Section VI.

## II. RELATED WORK

Network slicing has recently gained momentum among a growing community of academic and industrial researchers. Richart *et al.* [18] described how the resources of a wireless network can be appropriately allocated to network slices without having a detrimental impact on the quality of other slices. Foukas *et al.* [19] studied the notion of network slicing in 5G on the basis of an architectural model that consists of an infrastructure layer, a network function layer, and a service layer. Dawaliby *et al.* [20] formulated the problem of network slicing in a LoRa network as a one-to-many matching game and the problem of resource allocation within slices as a multiobjective optimization problem. Mai *et al.* [21] designed a network slicing architecture over an SDN-enabled LoRa network to dynamically divide the network into multiple virtual networks based on different service requirements. Heterogeneous wired and wireless networks can be partitioned into multiple slices via a network hypervisor to achieve isolation between the slices while sharing physical infrastructure, e.g., base stations, network switches, and links [22]. In [23], a network hypervisor named FlowVisor was proposed, which is commonly used in SDN to divide a physical network into several slices. Based on the FlowVisor hypervisor, Chen *et al.* [24] proposed a dynamic resource management platform called the EnterpriseVisor engine, which manages the distribution of network resources across slices, providing a basis for network slicing implementation. In [25] and [26], the connection admission control (CAC) mechanism was proposed as a dynamic network slice adjustment method. This protocol determines the acceptance or rejection of connection requests as well as their routing and allocation of network resources based on the remaining resources when devices access the network. As demonstrated in these existing studies, network slicing technology is critical to future communication networks, as it enables innovation and feature enhancement.

In the context of network slicing in power systems, current research primarily addresses the following three areas:

a) *Resource Allocation:* Hu et al. [27] introduced a one-dimensional convolutional neural network-based hierarchical scheduling system. This system intelligently classifies 5G network slicing for smart grid communication services, enabling self-driven resource allocation. An et al. [28] proposed a secure isolation technique for access-side communication services using network slicing in wireless local area network (WLAN). This technique allocates time-frequency resources to power grid communication services, constructing network slices for logical isolation and optimal use of communication resources. To enhance resource utilization efficiency and reduce operational costs for power companies, Zhong et al. [29] utilized the deep reinforcement learning ACKTR algorithm within an actor-critical framework. This approach determines the best policy for resource allocation and release in response to slice requests.

b) *QoS Assurance:* In [30], Kong et al. proposed a hybrid wireless access selection algorithm to ensure QoS in power services, achieving two-stage radio resource allocation in community smart grids. Zhao et al. [31] introduce a dynamic network slicing optimization scheme for 5G communication resources in smart grids, utilizing reinforcement learning to maintain QoS. A dynamic virtual network function scheduling strategy, based on a greedy algorithm for physical path selection and routing, was suggested in [32] to ensure sustainable QoS for power services. In [33], a resource scheduling and mapping algorithm using graph theory depth-first search (GDFS) was designed to meet the delay, rate, and reliability requirements of power services, aiming to reduce end-to-end delay. Li et al. [34] present a method for analyzing the trade-offs between radio access network (RAN) slice delay and throughput. This method, based on dynamic intelligent resource allocation, optimizes resource block allocation to ensure ultra-reliable and low latency communication (uRLLC) slice delays while meeting diverse QoS needs in smart grid environments.

c) *Network Resilience:* Carrillo et al. [35] proposed a novel RAN slicing framework utilizing artificial intelligence (AI) to support IEC 61850 smart grid services, aiming to meet self-healing performance requirements. In [36], enhanced network slicing, enabled by programmable hardware acceleration, isolates critical control and monitoring traffic in 5G networks for effective self-healing in smart grids. Another method, leveraging 5G communication and network slicing technologies, facilitates peer-to-peer communication between power distribution network nodes [37]. This approach enhances the fault isolation speed and minimizes the outage duration and frequency. The study in [13] proposed applying RAN slicing in 5G communication to assist virtual power plants (VPPs) in frequency regulation services. This enhances communication reliability, thereby ensuring regulation performance.

Although previous research has extensively explored network slicing in power systems from diverse perspectives, the primary focus has been on wireless network applications. This emphasis often overlooks the coexistence of optical fibers, cables, and wireless facilities in current PDS communication networks. Completely transitioning to wireless infrastructure is economically impractical. Therefore, effective utilization of existing wired and wireless communication technologies is essential in power systems to minimize equipment replacement costs. Furthermore, these studies overlook crucial network security enhancement challenges in power systems. Ensuring the cybersecurity of the PDS's communication network, the critical energy supplier for societal operations, is of utmost importance. Wireless networks, unlike their wired counterparts, are more susceptible to malicious attacks due to the broadcast nature of wireless communication [38], [39]. Indeed, some PDS network services use wired communication methods, primarily due to safety and security considerations. Therefore, these previous network slicing methods offer limited defense against cyberattacks in PDS communication networks. Moreover, implementing wireless technologies such as 5G, as suggested in these studies, might increase vulnerability of PDS to cyberattacks.

To tackle the mentioned challenges, we introduce a method based on network integration and authentication slice for network slicing. Diverging from prior studies, this method improves the utilization of existing network infrastructure and enhances network resilience by integrating multiple service networks and creating various slices according to service demands. Furthermore, this approach distinctively creates a dedicated authentication slice for device authentication, isolating newly connected devices within this slice rather than in the service slices. This method reduces the risk of unauthorized devices impersonating legitimate devices, illicitly accessing service data, and attacking servers and networks, thus bolstering network security. Furthermore, we introduce a network optimization method derived from our network slicing approach, which enables the PDS network to evolve with the progression of PDS. Consequently, our proposed method excels in utilizing existing facilities, providing network resilience, protection, and adaptability.

III. NETWORK SLICING FRAMEWORK AND MODEL

*A. Network Slicing Method*

PDS communications are mostly carried out in PDSs' own separate networks using dedicated physical cables, fibers, and devices [40], as shown in Fig. 2(a). While this dedicated and monolithic structural topology certainly provides an extra level of security, it provides no protection once an attacker has gained physical access to these networks [41]. Moreover, traditional PDS communication networks often neglect further strict security mechanisms such as cyberattack prevention, device authentication, or integrity checking [40]. To increase the network security of PDS, several researchers have proposed retrofitting terminal devices in traditional PDS networks with encryption chips [42], [43], thereby reducing the risk of illegitimate devices interacting with PDS operation centers. However, this method cannot protect a PDS communication network from illegitimate information acquisition, DOS, or FDI attacks.

To address the abovementioned problems, this paper proposes a method of optimizing traditional networks and allocating their communication resources by means of the following measures.

1) *Network Integration:* As shown in Fig. 2(b), we integrate multiple independent PDS communication networks into one network. Such network integration helps to realize communication resource sharing.
2) *Device Authentication:* We install encryption chips in terminal devices and authentication servers in PDS communication networks. Automatic device checking and identification are realized via authentication between the servers and terminal devices.
3) *Network Slicing:* As shown in Fig. 2(b), we divide this integrated network into slices in accordance with the communication requirements of the PDS services to be provided. Sharing of the same physical network infrastructure among different services under various scenarios is realized through network slice division.
4) *Authentication Slice*: In addition, a device authentication slice is specially established for the exchange of authentication information between terminal devices and authentication servers. The authentication slice includes all network ports, allowing servers to discover and authenticate devices. Devices that have not successfully completed authentication are restricted to sending and receiving information only in the authentication slice, thus achieving isolating of unauthorized, potentially risky behaviors.
5) *Dynamic Slice Division:* The ranges and bandwidths of various service slices are dynamically adjusted in accordance with real-time terminal device authentication results. In this way, device plug-and-play, automatic fault recovery and network risk isolation capabilities are achieved for the PDS communication networks.

For illustration, the process of device authentication is described for terminal device 8 in Fig. 2(c) as follows.

*i)* When terminal device 8 joins the PDS communication network, it sends its discovery information into the authentication slice for authentication. An authentication server then verifies the device information and access authority when it finds this new accessing device and receives its authentication information.
*ii)* If the server successfully authenticates this terminal device, it will pass the device access request to the network slice controller. The controller will then adjust the corresponding service slice to include terminal device 8 and allocate adequate communication resources in accordance with the service requirements.
*iii)* When terminal device 8 is connected to its service slice, it sends its identifier and model number to the operation center server for registration. After receiving this registration information, the server adjusts the database to establish data exchange with this terminal device.

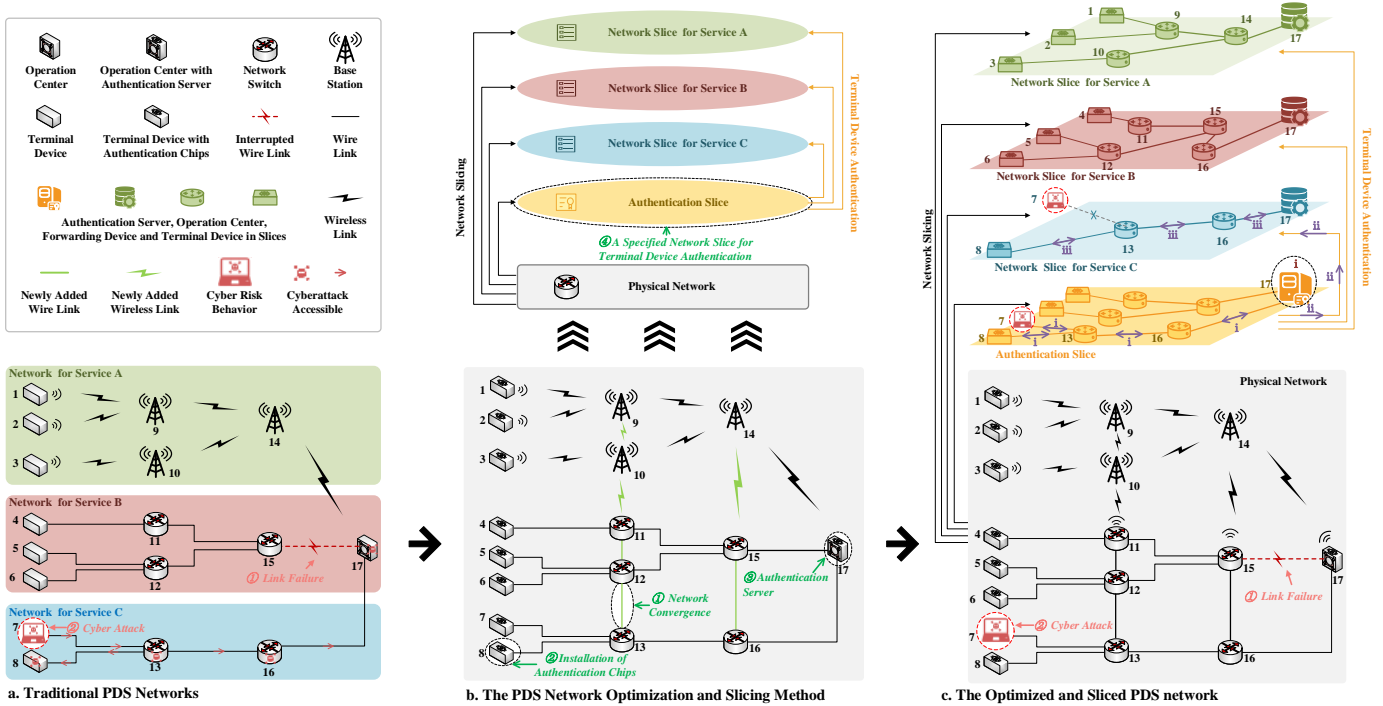

**Fig. 2.** Schematic illustration of the network slicing method.

The above approach enables fast access for terminal devices. Subsequently, the servers perform device verification with the terminal devices at regular time intervals. If a device fails to authenticate with the authentication servers, the server dynamically adjusts the associated slice range and removes the corresponding port from the server slice. This reduces the chance of information being stolen and other malicious attacks occurring on network slices. This method thereby enables a secure plug-and-play functionality for terminal devices.

The advantages of this method of network slicing can be illustrated through a comparison of the networks in Fig. 2(a) and Fig. 2(c) under the scenarios of network failure and network attack. First, if the link between network switch 15 and operation center 17 fails, the terminal devices of service B will all be disconnected from the operation center in Fig. 2(a), resulting in wide-area communication service interruption. In contrast, in the corresponding integrated network, these three PDS networks supporting independent communication services are interconnected to achieve the sharing of communication resources. As shown in Fig. 2(c), the remaining bandwidth resources of links, 15–16 and 16–17, and switches, 15 and 16, can be used to restore communication for service B, thus increasing the resilience of the network in coping with accidental failures. Second, consider the case in which an illegitimate device masquerades terminal device 7 and performs cyberattacks at the corresponding network interface. In the traditional network shown in Fig. 2(a), due to the lack of a rigorous device verification process, this illegitimate device can easily disguise itself as a legitimate device to obtain various types of PDS operational and control data and can even send malicious control commands to PDS facilities to trigger a physical layer failure. In addition, the above device is able to carry out DOS attacks on the operation center server or inject a large amount of false data through the port to disrupt the normal network communication. The direct application of common network slicing schemes does not help to reduce this type of risk. However, in our device authentication slice-based sliced network shown in Fig. 2(c), newly accessing devices are restricted to the authentication slice and are rigorously verified. On the one hand, it is difficult for illegitimate devices to pass the device encryption-based authentication process; thus, these devices can be isolated and prevented from obtaining PDS operational data. On the other hand, when a DOS or FDI attack is conducted through port 7, the attack is restricted to the device authentication slice and thus has difficulty affecting normal communication services. When the authentication server detects a cyber risk at this port, it removes the corresponding connection in the authentication slice and waits for manual detection to resume. This approach greatly enhances the security of the PDS communication network.

In addition, considering that the use of a single centralized authentication server may make the entire network susceptible to authentication server failures caused by malicious attacks or equipment malfunctions, we propose installing multiple decentralized authentication servers in the network, subject to cost constraints. This approach reduces the influence of the failure of a single authentication server, and thus improves the operational safety of the PDS.

The communication network is divided into various communication network slices based on the associated PDS communication services, as described in the following steps,

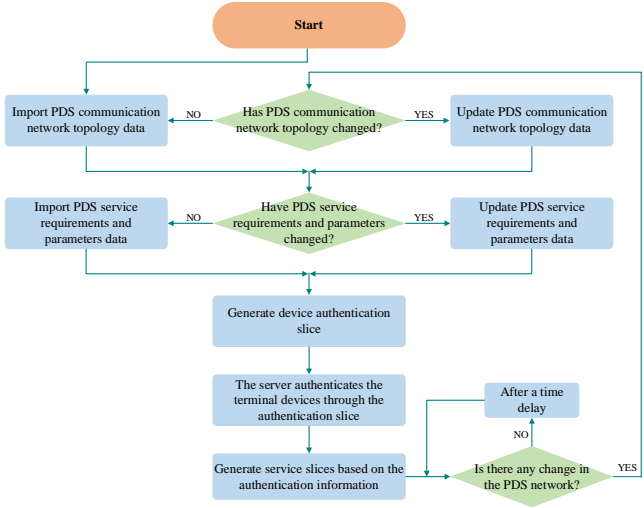

**Fig. 3.** Flowchart of the network slicing methodology.

and illustrated in Fig. 3:

1) In the communication network, a basic authentication slice is first created to establish communication between the authentication servers and as many ports in the network as possible for device authentication.
2) When terminal devices access the network, if an authentication server is successful at authenticating these devices, then the corresponding service slices will be adjusted to include appropriate interface ports. All slices are customized in accordance with the range of the accessing devices, service requirements, and service priority.
3) The authentication server polls the network devices at fixed intervals. When a device exits the network, the authentication servers remove its interface port from each service slice to prevent unauthorized access. The network slices are adjusted accordingly when interfaces with new devices are established.

This network slicing scheme ensures the efficient, dynamic, and reasonable allocation of communication resources. When communication resources are limited, resources are preferentially allocated to critical communication services to protect important service operations. Moreover, this device authentication slice-based network slicing scheme protects network services from illegitimate access.

Some limitations should be acknowledged. Initially, despite enhancing the use of existing communication link facilities, this method necessitates modifying forwarding equipment to facilitate network slicing. Secondly, although the authentication slice and server-based approach mitigate various cyberattacks, the vulnerabilities of authentication server to DoS attacks persist. We contend that authentication servers, as specialized network security devices, possess DoS attack countermeasures, rendering them less susceptible to DoS and distributed DoS (DDoS) attacks. Moreover, to lessen the impact of a single authentication server's failure from DoS attacks, we advocate for a decentralized authentication server deployment strategy, recognizing that achieving total defense remains an ongoing objective. Future research endeavors will concentrate on overcoming these limitations.

*B. Network Slicing Model*

From the topological perspective, the PDS communication network can be represented as an undirected graph $\mathcal{G} = (\mathcal{N}, \mathcal{L})$ with $N_n$ network nodes interconnected by $N_l$ network links. Moreover, the node set $\mathcal{N}$ can be subdivided into a set of terminal nodes (e.g., terminal devices) $\mathcal{N}^e \subset \mathcal{N}$, a set of forwarding nodes (e.g., network switches and base stations), $\mathcal{N}^f \subset \mathcal{N}$, and a set of nodes with authentication servers, $\mathcal{N}^a \subset \mathcal{N}$. Due to the performance limitations of the actual equipment, these nodes and links have bandwidth limitations. Each forwarding node introduces a delay in the forwarding process, and each link introduces a propagation delay in the transmission process. The types of services, along with their delay and bandwidth requirements, are specified in a subset $\mathcal{B}$.

*1) Slice Accessing Constraints:* Nodes or links in the communication network may fail due to extreme events or malicious damage. Only normally functioning nodes can interface with the network slices, i.e.,

$$n_i^x \leq \varsigma_i^N, \forall i \in \mathcal{N}, \forall x \in \mathcal{B}, \quad (1)$$
$$l_k^x \leq \varsigma_k^L, \forall k \in \mathcal{L}, \forall x \in \mathcal{B}, \quad (2)$$

where $l_k^x$ and $n_i^x$ denote whether link $k$ and node $i$, respectively, are interfaced into slice $x$, with "1" indicating interfaced and "0" indicating not interfaced; similarly, $\varsigma_k^L$ and $\varsigma_i^N$ denote the working states of link $k$ and node $i$, respectively, with "1" indicating normal and "0" indicating faulty.

Considering the security of various transmission methods (for example, some service data with high security requirements cannot be transmitted on wireless paths), we introduce positive integer parameters $\theta_i^n$ and $\theta_k^l$ to represent the security levels of node $i$ and link $k$, respectively. A positive integer parameter $\vartheta^x$ is used to denote the security level required for service $x$ [44], [45]. The data for service $x$ can be transmitted only on links and nodes that meet the corresponding security requirement; this restriction can be formulated as the following two inequality constraints:

$$n_i^x \leq \theta_i^n / \vartheta^x, \forall i \in \mathcal{N}, \forall x \in \mathcal{B}\backslash a, \quad (3)$$
$$l_k^x \leq \theta_k^l / \vartheta^x, \forall k \in \mathcal{L}, \forall x \in \mathcal{B}\backslash a, \quad (4)$$

where service "$a$" represents the authentication service.

The network topology must also be radially connected to prevent data from circulating, and this constraint can be expressed as follows:

$$\sum_{k \in \mathcal{L}} l_k^x = \sum_{i \in \mathcal{N}} n_i^x - 1, \forall x \in \mathcal{B}\backslash a, \quad (5)$$
$$\sum_{k \in \mathcal{L}} l_k^a = \sum_{i \in \mathcal{N}} n_i^a - \sum_{i \in \mathcal{N}^a} n_i^a. \quad (6)$$

A connection through the authentication slice must be established between the terminal devices and the authentication server to allow a device to access the corresponding service slice; thus, we have:

$$n_i^x \leq s_i^x, \forall i \in \mathcal{N}^e, \forall x \in \mathcal{B}\backslash a, \quad (7)$$

where $s_i^x$ represents the authentication state of node $i$ for service $x$, with "1" indicating successful authentication and "0" indicating unsuccessful authentication.

For the operation center node, the data to be received should consist of all the data on the connected communication links. For forwarding nodes, there should be two procedures for sending and receiving identical data. For terminal nodes, there should be only one data transmission process. The above characteristics can be translated into the following three linear equality constraints:

$$\sum_{k \in \mathcal{F}_o} f_k^x = g_o^x, \forall x \in \mathcal{B}, \tag{8}$$

$$\sum_{k \in \mathcal{F}_i} f_k^x = 2g_i^x, \forall x \in \mathcal{B}, \forall i \in \mathcal{N}^f, \tag{9}$$

$$\sum_{k \in \mathcal{F}_i} f_k^x = E_y(i,:) \cdot n_i^x, \forall x \in \mathcal{B}, \forall i \in \mathcal{N}^e, \tag{10}$$

where node $o$ represents the operation center; $f_k^x = (b_1^{L,k,x}, \ldots, b_y^{L,k,x})$ and $g_i^x = (b_1^{N,i,x}, \ldots, b_y^{N,i,x})$ are two vectors consisting of binary variables, with $b_j^{L,k,x}$ and $b_j^{n,i,x}$ being binary variables indicating whether the data of terminal node $j$ for service $x$ flow through link $k$ and node $i$, respectively; $\mathcal{F}_i$ is the set of links connected to node $i$; and $E_y(i,:)$ denotes row $i$ of the $y$-dimensional identity matrix.

The data for each type of service can flow only on the nodes and links that belong to the corresponding slice, which can be expressed as follows:

$$b_j^{L,k,x} \leq l_k^x, j \in \mathcal{N}^e, \forall x \in \mathcal{B}, \forall k \in \mathcal{L}, \tag{11}$$

$$b_j^{N,i,x} \leq n_i^x, j \in \mathcal{N}^e, \forall x \in \mathcal{B}, \forall i \in \mathcal{N}^f. \tag{12}$$

The single-commodity flow approach is used to guarantee the connectivity constraints of the nodes in each slice with respect to the operation center node. Thus, we can express these connectivity constraints as follows:

$$\sum_{k \in \mathcal{F}_o} m_k^x \cdot \mu_{o,k} - n_o^x = h_o^x, \forall x \in \mathcal{B}, \tag{13}$$

$$\sum_{k \in \mathcal{F}_i} m_k^x \cdot \mu_{i,k} - n_i^x = 0, x \in \mathcal{B}, \forall i \in (\mathcal{N}^f \cup \mathcal{N}^e). \tag{14}$$

Here, $m_k^x$ denotes the commodity flow from node $i$ to node $j$ on link $k = (i,j)$. $\mu_{i,k}$ is the assumed commodity flow direction parameter on link $k$ with respect to node $i$; specifically, the flow direction on link $k = (i,j)$ is defined as the direction from node $i$ to node $j$, as expressed by parameters $\mu_{i,k} = 1$ and $\mu_{j,k} = -1$. $h_o^x$ is the output commodity flow of the operation center node.

To ensure the connectivity of the slices, the interface states of the nodes at both ends of a connected link on a service slice are constrained to be the same. Accordingly, the node relationships can be described as follows:

$$-(1 - l_k^x) + n_j^x \leq n_i^x \leq n_j^x + (1 - l_k^x), \forall k = (i,j) \in \mathcal{L}. \tag{15}$$

*2) Bandwidth and Delay Constraints:* The bandwidth consumption of a node or link is equal to the sum of the bandwidth consumption of the data flowing through it and cannot exceed its bandwidth capacity. These constraints can be expressed as

$$w_k^{L,x} = \sum_{b_j^{L,k,x} \in f_k^x} b_j^{L,k,x} \cdot c_j^x, \forall x \in \mathcal{B}, \forall k \in \mathcal{L}, \tag{16}$$

$$\sum_{x \in \mathcal{B}} w_k^{L,x} \leq \lambda_k^L, \forall k \in \mathcal{L}, \tag{17}$$

$$w_i^{N,x} = \sum_{b_j^{N,i,x} \in g_i^x} b_j^{N,i,x} \cdot c_j^x, \forall x \in \mathcal{B}, \forall i \in \mathcal{N}^f, \tag{18}$$

$$\sum_{x \in \mathcal{B}} w_i^{N,x} \leq \lambda_i^N, \forall i \in \mathcal{N}^f, \tag{19}$$

where $w_k^{L,x}$ and $w_i^{N,x}$ denote the bandwidths consumed by the data for service $x$ on link $k$ and node $i$, respectively; $c_j^x$ is the normal required communication bandwidth for terminal node $j$ for service $x$; and $\lambda_k^L$ and $\lambda_i^N$ denote the upper bandwidth limits of link $k$ and node $i$, respectively.

The end-to-end communication delay between a node and the operation center is the sum of the forwarding delays of the forwarding nodes and the propagation delays of the links through which the data are passed. This end-to-end delay cannot exceed the specified delay limitation, that is,

$$d_i^x \leq d_{max}^x, \forall i \in \mathcal{N}, \forall x \in \mathcal{B}, \tag{20}$$

$$d_j^x = \sum_{k \in \mathcal{L}} b_j^{L,k,x} \cdot \xi_k^L + \sum_{i \in \mathcal{N}^f} b_j^{N,i,x} \cdot \xi_i^N, \forall j \in \mathcal{N}^e, \forall x \in \mathcal{B}, \tag{21}$$

where $d_i^x$ is the end-to-end data delay from node $i$ to the operation center; $d_{max}^x$ denotes the upper limit on the end-to-end delay for service $x$; and $\xi_k^L$ and $\xi_i^N$ denote the propagation delay of link $k$ and the forwarding delay of node $i$, respectively.

*3) Objective Function:* The goal of the whole network slicing process is to achieve the maximum value of service connections under the various constraints given above, which can be expressed as a two-stage optimization problem. The goal of the first stage is to solve for the authentication slice topology with the maximum number of connected nodes, for which the objective function can be expressed as

$$max \sum_{i \in \mathcal{N}^e} n_i^a, \tag{22}$$

$$s.t. (1) - (4), (6), (8) - (21).$$

The results of the first-stage model are the basis for the second-stage model. After terminal device authentication, the goal of the second stage is to solve for the maximum service value of the network. The second-stage objective function is

$$max \sum_{x \in \mathcal{B}/a} \gamma^x \cdot \sum_{i \in \mathcal{N}^e} n_i^x, \tag{23}$$

$$s.t. (1) - (5), (7) - (21).$$

This problem is formulated as an MILP model, which can be effectively solved by off-the-shelf commercial solvers.

## IV. NETWORK ROBUSTNESS EVALUATION AND OPTIMIZATION

### A. Network Robustness Evaluation

Network robustness is defined as the ability of a network to withstand failures [46], and is an important metric for evaluating a network's resistance to arbitrary disruptions such as equipment failures, natural disasters, and malicious attacks [39]. Therefore, we propose a robustness evaluation method based on the nodes and links in a PDS communication network. This robustness evaluation method assumes failures in each link and each node in the network and evaluates the robustness with respect to each component based on the resulting service value after its failure. Notably, the network performance is significantly affected by the links between forwarding nodes as well as the links between forwarding nodes and the operation center. Thus, the robustness evaluation, robustness optimization, and load balancing regulation processes presented in this paper all focus on these links and nodes.

We perform a robustness evaluation of a normally functioning communication network by invoking the above network slicing model. Since the terminal devices that need to access the network are known at this point, the objective function for the robustness evaluation can be simplified to the objective function of the second-stage problem, subject to the constraints of the authentication slice. Accordingly, we define two functions, $u_z^L(\eta)$ and $u_y^N(\eta)$, to help us evaluate the robustness of the links and nodes:

$$u_z^L(\eta) = max \sum_{x \in \mathcal{B} \setminus a} \gamma^x \cdot \sum_{i \in \mathcal{N}^e} n_i^x \quad (24)$$
$$s.t. (1) - (21),$$
$$\varsigma_z^L = \eta, \varsigma_k^L = 1, \varsigma_i^N = 1, \forall k \in \mathcal{L} \setminus z, \forall i \in \mathcal{N}^f, \quad (25)$$
$$u_y^N(\eta) = max \sum_{x \in \mathcal{B} \setminus a} \gamma^x \cdot \sum_{i \in \mathcal{N}^e} n_i^x \quad (26)$$
$$s.t. (1) - (21),$$
$$\varsigma_y^N = \eta, \varsigma_i^N = 1, \varsigma_k^L = 1, \forall i \in \mathcal{N}^f \setminus y, \forall k \in \mathcal{L}. \quad (27)$$

Here, $\eta$ is a parameter taking a value of 0 or 1, and $u_z^L(\eta)$ and $u_y^N(\eta)$ are functions designed to solve for the maximum value of service connections under the constraints that the working states of link $z$ and node $y$, respectively, are normal or faulty.

Then, the link and node robustness evaluation functions can be defined as follows:

$$p_z^L = u_z^L(1) - u_z^L(0), \forall z \in \mathcal{L}, \quad (28)$$
$$p_y^N = u_y^N(1) - u_y^N(0), \forall y \in \mathcal{N}^f, \quad (29)$$
$$r_z^L = \omega_z^L - p_z^L, \forall z \in \mathcal{L}, \quad (30)$$
$$r_y^N = \omega_y^N - p_y^N, \forall y \in \mathcal{N}^f, \quad (31)$$

where $p_z^L$ and $p_y^N$ denote the effect-on-service values when link $z$ and node $y$, respectively, fail; $\omega_z^L$ and $\omega_y^N$ denote the fault resistance ability (FRA) parameters of link $z$ and node $y$, respectively, which are related to the transmission mode and equipment type; and $r_z^L$ and $r_y^N$ denote the robustness evaluation functions for link $z$ and node $y$, respectively. The robustness evaluation method can be formulated as shown in Algorithm 1:

---

**Algorithm 1** Network robustness evaluation

**Input** the whole network topology $\mathcal{G} = \{\mathcal{N}, \mathcal{L}\}$ along with its delay, bandwidth, security, and FRA parameters;
**Input** the network services along with their delay, bandwidth, and security requirements;

1: **function** $NetworkSlicing(\{\varsigma_i^N, \forall i \in \mathcal{N}\}, \{\varsigma_k^L, \forall k \in \mathcal{L}\})$
2:     Compute the network slicing solution with the maximum service value $u$ in the above network slicing model given the link working states $\{\varsigma_i^N, \forall i \in \mathcal{N}\}$ and node working states $\{\varsigma_k^L, \forall k \in \mathcal{L}\}$;
3:     **output** $u$;
4: Calculate the service value when the nodes and links are normal: $u^a = NetworkSlicing(\{\varsigma_i^N = 1, \forall i \in \mathcal{N}\}, \{\varsigma_k^L = 1, \forall k \in \mathcal{L}\})$;
5: **for** each $z \in \mathcal{L}$ **do**
6:     Calculate the maximum service value when link $z$ fails: $u_z^L = NetworkSlicing(\{\varsigma_i^N = 1, \forall i \in \mathcal{N}\}, \{\varsigma_z^L = 0, \varsigma_k^L = 1, \forall k \in \mathcal{L} \setminus z\})$;
7:     Compute and save the effect-on-service value $p_z^L = u^a - u_z^L$, then compute the robustness evaluation value $r_z^L = \omega_z^L - p_z^L$;
8: **end for**
9: **for** each $y \in \mathcal{N}^f$ **do**
10:    Calculate the maximum service value when forwarding node $y$ fails: $u_y^N = NetworkSlicing(\{\varsigma_y^N = 0, \varsigma_i^N = 1, \forall i \in \mathcal{N} \setminus y\}, \{\varsigma_k^L = 1, \forall k \in \mathcal{L}\})$;
11:    Compute and save the effect-on-service value $p_y^N = u^a - u_y^N$, then compute the robustness evaluation value $r_y^N = \omega_y^N - p_y^N$;
12: **end for**
**Output** the robustness evaluation results: $\{r_y^N, \forall y \in \mathcal{N}^f\}$ and $\{r_z^L, \forall z \in \mathcal{L}\}$.

---

### B. Network Robustness Optimization

A growing number of participating terminal devices is causing the utilization rates of the communication network equipment and links to progressively increase. Consequently, an increasing number of services will be affected when network components fail, meaning that the robustness with respect to various network components is simultaneously reduced. In this scenario, optimizing the network topology to reduce the load rates on network components and improve the network robustness is crucial for PDS communication networks.

Based on the robustness evaluation functions for the different types of network components described above, we propose a network optimization method. While minimizing the cost of robustness optimization, the method derives a network optimization scheme that satisfies the robustness and load balancing requirements of the network by optimizing the links and nodes. The optimization of links includes the upgrading of original links and the addition of new links, while the optimization of nodes includes the upgrading and replacement of forwarding equipment. Stacking technology [47] is considered helpful for upgrading forwarding devices. The optimization scheme is implemented to improve the robustness and load balance of the network.

*1) Optimization Cost Constraints:* Three variables, $\tau_k^{L,a}$, $\tau_k^{L,m}$, and $\tau_i^{N,m}$, are introduced to represent the costs of adding a link $k$, optimizing a link $k$, and optimizing a forwarding node $i$, respectively. The cost of optimizing a link should be related to the length of the link and the geographical factors in the region through which the link passes, whereas the cost of optimizing a node is related only to the cost of the equipment used for optimization. Accordingly, the optimization costs are formulated as follows:

$$\tau_k^{L,a} = \alpha_{ij} + \varphi_y^{L,a} \cdot \varpi_{ij}, \forall k = (i,j) \in \mathcal{L}^a, \forall y \in \mathcal{F}_k^{L,m}, \quad (32)$$
$$\tau_k^{L,m} = \beta_{ij} + \varphi_y^{L,m} \cdot \varpi_{ij}, \forall k = (i,j) \in \mathcal{L}, \forall y \in \mathcal{F}_k^{L,a} \quad (33)$$
$$\tau_i^{N,m} = \beta_y^N, \forall k = (i,j) \in \mathcal{L}, \forall y \in \mathcal{F}_i^{N,m}, \quad (34)$$

where $\varpi_{ij}$ is the distance between nodes $i$ and $j$; $\alpha_{ij}$ and $\beta_{ij}$ are the geographical condition-related construction and optimization costs, respectively, for a link $k$ between nodes $i$ and $j$; $\varphi_y^{L,a}$ and $\varphi_y^{L,m}$ are the average per-length construction and optimization costs, respectively, for link $k$ when construction/optimization option $y$ is selected; $\mathcal{L}^a$ is the set of possible links that could be added; and $\beta_y^N$ represents the cost of option $y$ for optimizing forwarding node $i$.

*2) Optimization Option Constraints:* A binary variable $q_{k,y}^{L,a}$ is introduced to indicate whether construction plan $y$ is chosen for link $k$. Binary variables $q_{k,y}^{L,m}$ and $q_{k,y}^{L,m}$ are introduced to indicate whether optimization option $y$ is chosen for link $k$ and node $i$, respectively. Note that the optimization of links and nodes involves choices among several options, which can be formulated as three inequality constraints:

$$\sum_{y \in \mathcal{F}_k^{L,a}} q_{k,y}^{L,a} \leq 1, \forall k \in \mathcal{L}^a, \quad (35)$$
$$\sum_{y \in \mathcal{F}_k^{L,m}} q_{k,y}^{L,m} \leq 1, \forall k \in \mathcal{L}, \quad (36)$$
$$\sum_{y \in \mathcal{F}_i^{N,m}} q_{i,y}^{N,m} \leq 1, \forall i \in \mathcal{N}^f, \quad (37)$$

where $\mathcal{F}_k^{L,a}$, $\mathcal{F}_k^{L,m}$, and $\mathcal{F}_i^{N,m}$ are the sets of link construction options, link optimization options, and node optimization options, respectively.

*3) Bandwidth Constraints:* In the presence of multiple

optimization options, the possible bandwidths of the various links and nodes are:

$$\lambda_k^L = \lambda_k^{L,p} + \sum_{y \in \mathcal{F}_k^{L,m}} q_{k,y}^{L,m} \cdot \lambda_{k,y}^{L,m}, \forall k \in \mathcal{L}, \quad (38)$$

$$\lambda_k^L = \sum_{y \in \mathcal{F}_k^{L,a}} q_{k,y}^{L,a} \cdot \lambda_{k,y}^{L,a}, \forall k \in \mathcal{L}^a, \quad (39)$$

$$\lambda_i^N = \lambda_i^{N,p} + \sum_{y \in \mathcal{F}_i^{N,m}} q_i^{N,m} \cdot \lambda_{i,y}^{N,m}, \forall i \in \mathcal{N}^f, \quad (40)$$

where $\lambda_k^{L,p}$ and $\lambda_i^{N,p}$ are the bandwidths of link $k$ and node $i$, respectively, before the optimization procedure; $\lambda_k^{L,m}$ and $\lambda_i^{N,m}$ denote the increased bandwidths of link $k$ and node $i$, respectively, when optimization option $y$ is applied; and $\lambda_{k,y}^{L,a}$ denotes the bandwidth of link $k$ that is added when optimization option $y$ is chosen.

To reduce the load rates on the switches and links, we first denote the bandwidth utilization rates of the links and switches by $\chi^L$ and $\chi^N$, respectively. Then, the bandwidth utilization rates of the nodes and links must satisfy the following constraints:

$$\sum_{x \in \mathcal{B}} w_k^{L,x} \leq \chi^L \cdot \lambda_k^L, \forall k \in (\mathcal{L}^a \cup \mathcal{L}), \quad (41)$$

$$\sum_{x \in \mathcal{B}} w_i^{N,x} \leq \chi^N \cdot \lambda_i^N, \forall i \in \mathcal{N}, \quad (42)$$

*4) Robustness Constraints:* The optimization model is based on the original link and node robustness assessment formulas, which constrain the service data passing through the links and nodes. To guarantee sufficient robustness, it is specified that the weight of the FRA parameter for each link or node minus the total value of services passing through that link or node in the optimization model cannot reduce to below a certain lower bound. Moreover, the robustness with respect to the various links and nodes is calculated while accounting for the existence of multiple optimization options, and the corresponding equations are defined as follows:

$$\omega_k^L = \omega_k^{L,p} + \sum_{x \in \mathcal{F}_k^{L,m}} q_k^{L,m} \cdot \omega_k^{L,m}, \forall k \in \mathcal{L}, \quad (43)$$

$$\omega_k^L = \sum_{x \in \mathcal{F}_k^{L,a}} q_k^{L,a} \cdot \omega_k^{L,a}, \forall k \in \mathcal{L}^a, \quad (44)$$

$$\omega_i^N = \omega_i^{N,p} + \sum_{x \in \mathcal{F}_i^{N,m}} q_i^{N,m} \cdot \omega_i^{N,m}, \forall i \in \mathcal{N}^f, \quad (45)$$

$$\omega_k^L - \sum_{x \in \mathcal{B}} \gamma^x \cdot \sum_{i \in \mathcal{N}^e} b_i^{L,k,x} \geq \mu^L, \forall k \in (\mathcal{L}^a \cup \mathcal{L}), \quad (46)$$

$$\omega_y^N - \sum_{x \in \mathcal{B}} \gamma^x \cdot \sum_{i \in \mathcal{N}^e} b_i^{N,y,x} \geq \mu^N, \forall y \in \mathcal{N}, \quad (47)$$

where $\mu^L$ and $\mu^N$ are the robustness criteria for nodes and links, respectively; $\omega_k^{L,p}$ and $\omega_i^{N,p}$ are the FRA parameters of link $k$ and node $i$, respectively, before the optimization procedure; and $\omega_k^{L,m}$ and $\omega_i^{N,m}$ denote the increased FRA parameters of link $k$ and node $i$, respectively, when optimization option $y$ is applied.

*5) Terminal Node Access Constraints:* The original normal communication nodes associated with each service in the normal network are constrained to be able to communicate with the operation center in the optimized model. The related constraints can be written as

$$n_i^x = 1, \forall i \in \mathcal{A}^x, \forall x \in \mathcal{B}, \quad (48)$$

$$n_i^x = 0, \forall i \in \mathcal{N}^e \backslash \mathcal{A}^x, \forall x \in \mathcal{B}, \quad (49)$$

where $\mathcal{A}^x$ is the set of terminal device nodes that need to access service $x$ in the normally operating network.

*6) Objective Function:* The goal of the robustness optimization model is to find an optimization plan with the minimum cost that satisfies the load balancing constraints as well as the robustness requirements. The objective function can be formulated as follows:

$$\min(\sum_{k \in \mathcal{L}^a} \tau_k^{L,a} \cdot q_k^{L,a} + \sum_{k \in \mathcal{L}} \tau_k^{L,m} \cdot q_k^{L,m} + \sum_{i \in \mathcal{N}} \tau_i^{N,m} \cdot q_i^{N,m}),$$
$$(50)$$
$$s.t. (1) - (16), (18), (20) - (21), (32) - (49).$$

Then, the optimal scheme for reducing the network load rate while improving the network robustness can be obtained by solving the above model.

## V. NUMERICAL RESULTS

### A. Verification Cases for the Network Slicing Method

To verify the effectiveness of our proposed scheme, we consider a specific PDS communication network consisting of 60 nodes. This network includes 56 terminal devices, 10 network switches, 3 base stations, and 1 operation center, as depicted in Fig. 4. It supports 4 PDS services, each with its own communication network (color-coded as blue, red, green, or orange in Fig. 4): A. renewable power generation control; B. feeder automation; C. electric vehicle charging; and D. power equipment monitoring. The link bandwidth, link propagation delay, node bandwidth, and node forwarding delay parameters are also shown in Fig. 4. In addition, the authentication bandwidth requirement for each node is set to 0.5 kbps.

In accordance with the classification of service importance in a PDS [45], the bandwidth requirements, delay requirements, and weight coefficients of these four services are set as shown in Table I.

TABLE I
REQUIREMENTS AND PARAMETERS OF DIFFERENT SERVICES

| | Service | Bandwidth requirement | Delay requirement | Security requirement | Weight value |
|---|---|---|---|---|---|
| A. | renewable power generation control | 4 Mbps | 20 ms | 3 | 1.0 |
| B. | feeder automation | 2 Mbps | 20 ms | 2 | 0.8 |
| C. | electric vehicle charging | 0.4 Mbps | 1 s | 1 | 0.2 |
| D. | power equipment monitoring | 0.2 Mbps | 3 s | 2 | 0.4 |

To validate the slicing scheme proposed in this paper, three network cases are considered for comparison.

*a) Networks without integration and slicing:* The traditional PDS communication networks.

*b) Networks with integration and conventional slicing method:* An integrated PDS communication network sliced by means of an ordinary CAC mechanism-based network slicing scheme [25].

*c) Networks with integration and the proposed slicing method:* An integrated PDS communication network sliced by means of the proposed authentication slice-based network slicing method.

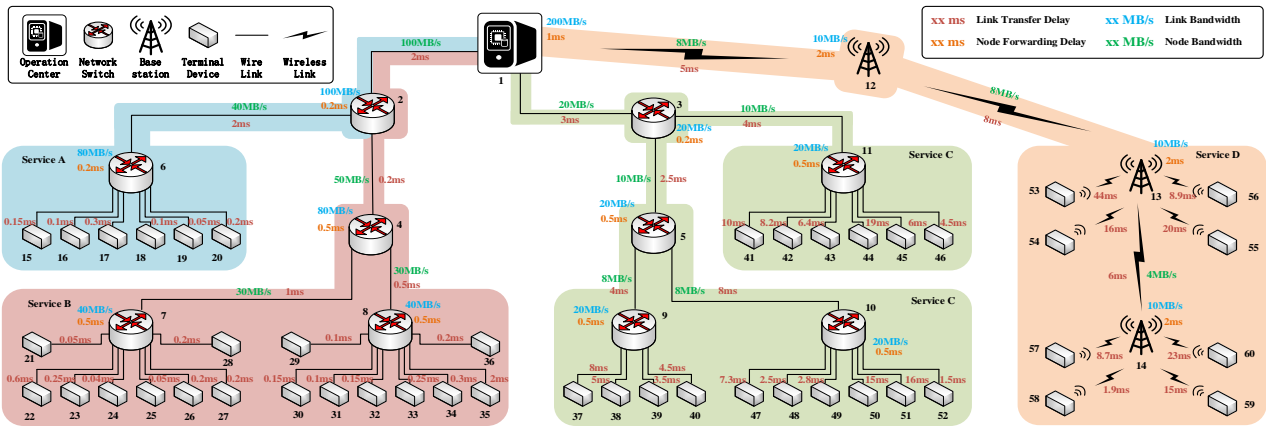

**Fig. 4.** Traditional PDS communication network topology case.

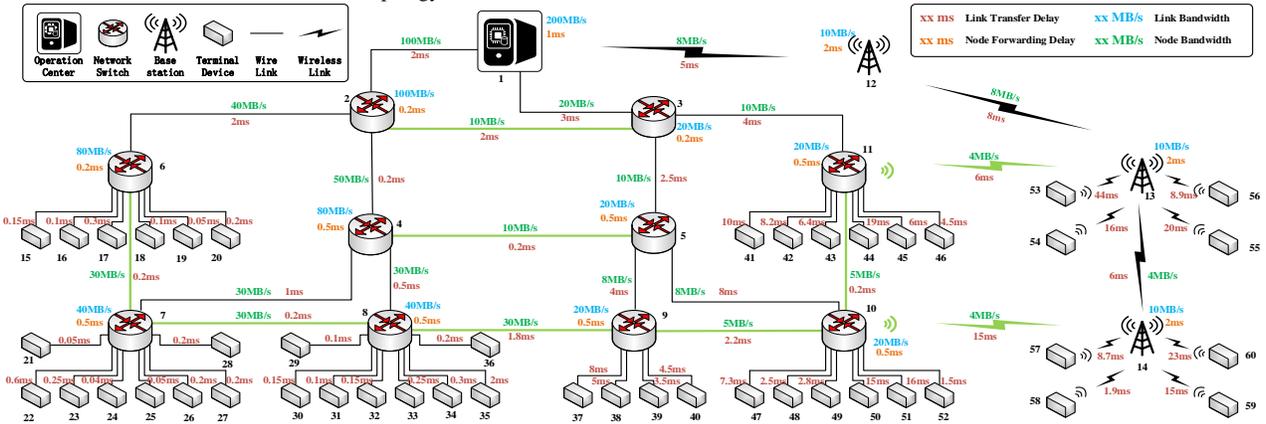

**Fig. 5.** Integrated PDS communication network topology case.

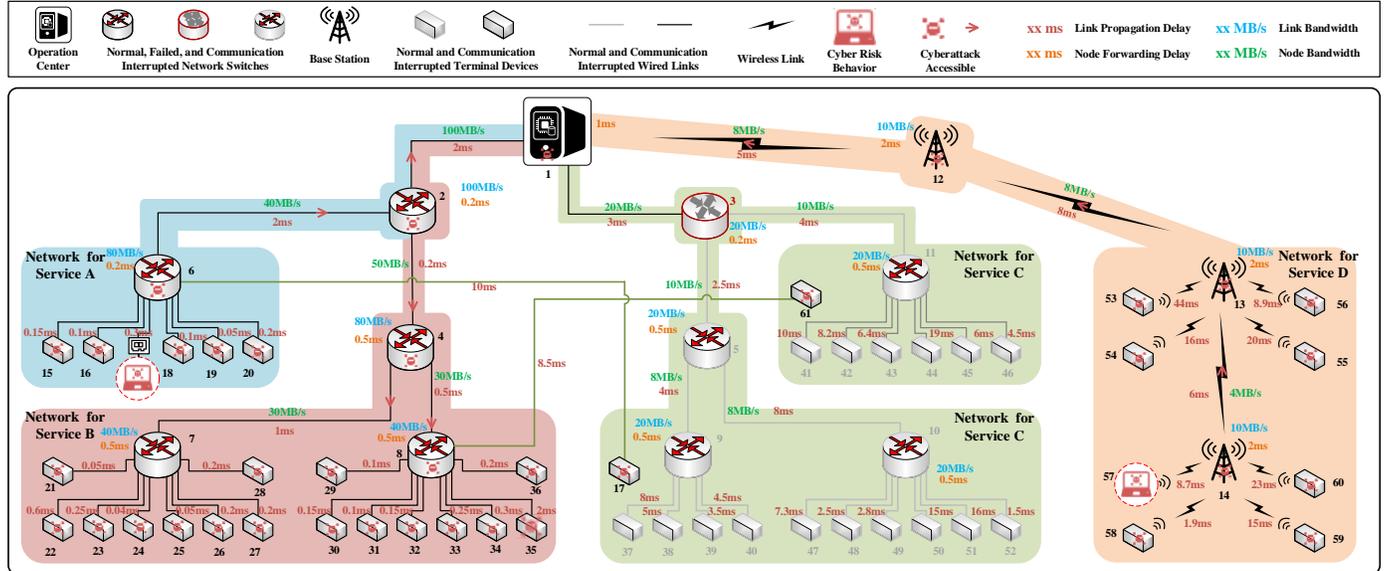

**Fig. 6.** Traditional networks in this scenario.

In accordance with the adjacency relationships in physical space, additional links have been artificially introduced among forward devices to facilitate network integration and bandwidth resource sharing. The security level parameters for the original nodes and links are detailed in Table II. Table IV lists these additional links with their parameters, and Fig. 5 presents a diagram of the network structure, highlighting the added links (both wired and wireless) in green.

TABLE II
SECURITY LEVEL PARAMETERS OF NODES AND LINKS

| Security level | Node | Link |
|---|---|---|
| 5 | 1–3 | 1–2, 1-3 |
| 4 | 6, 15–20, 61 | 2–6, 2-4 |
| 3 | 4–5, 7–12, 21–66 | 4–7, 4–8, 3–5, 5–9, 5–10, 3–11,1–12 |
| 2 | 13–14, 43–60 | 12–13, 13–14 |

Now, the network performance is tested considering the following problems in the overall PDS communication network:

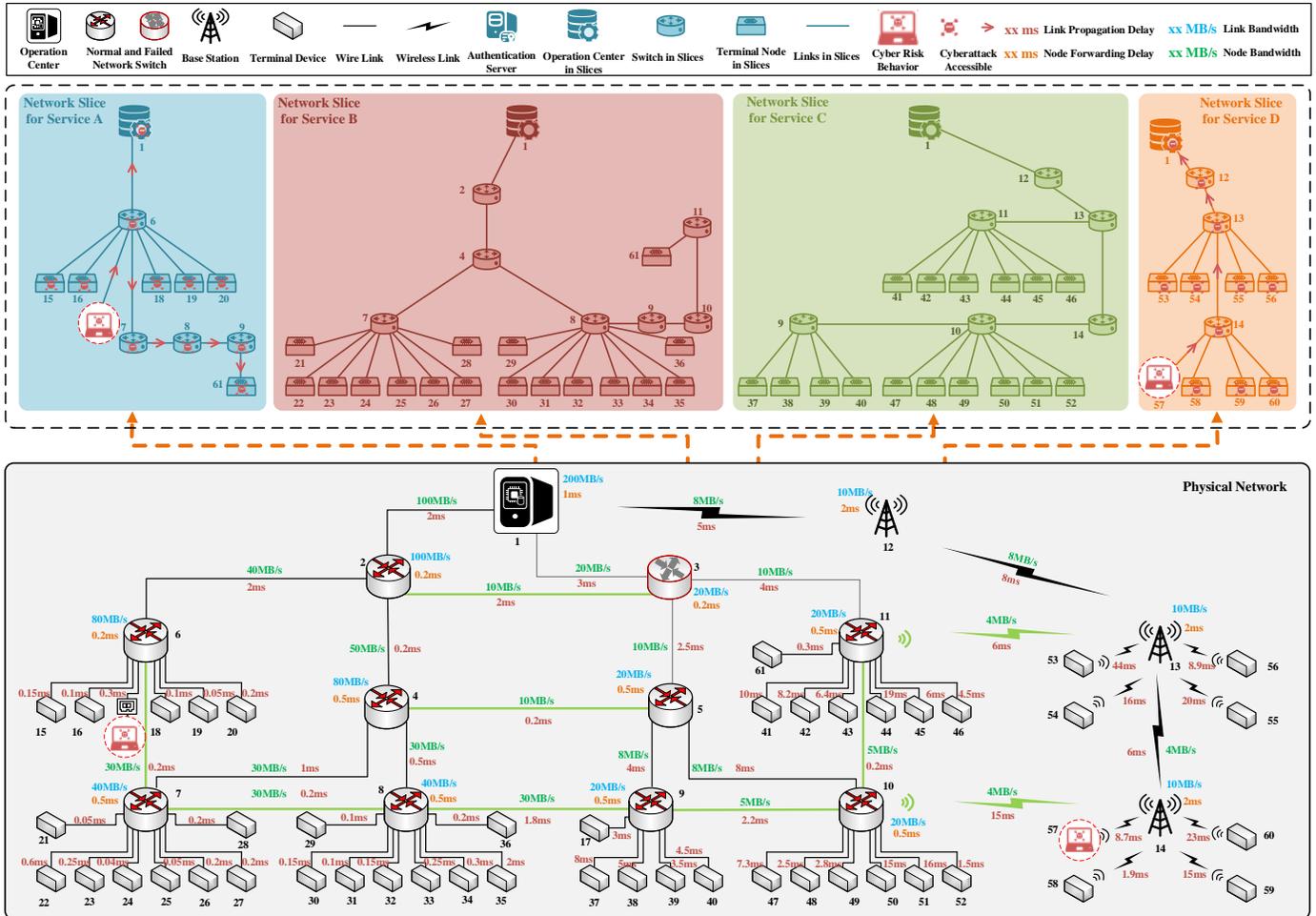

**Fig. 7.** The integrated network sliced by means of the CAC mechanism-based network slicing method in this scenario.

1) *Equipment failure:* Switch 3 fails in the PDS communication network above.
2) *Addition of equipment:* With the development of the PDS, a new terminal device 61 near switch 11 needs to be provided access to service B.
3) *Equipment relocation:* Due to the PDS service adjustment, the equipment at node 17 is moved near switch 9 and must be reconnected to service A.

Now, the network performance is tested considering the following problems in the overall PDS communication network:

1) *Equipment failure:* Switch 3 fails in the PDS communication network above.
2) *Addition of equipment:* With the development of the PDS, a new terminal device 61 near switch 11 needs to be provided access to service B.
3) *Equipment relocation:* Due to the PDS service adjustment, the equipment at node 17 is moved near switch 9 and must be reconnected to service A.
4) *Cyberattack:* The remaining network port of the removed terminal device 17 and the network link of terminal device 57 are illegitimately used for information stealing and network attacks.

The calculations for the three different network cases are implemented on a PC with an Intel Core i7 CPU @2.90 GHz and 32 GB of memory. The YALMIP toolbox in MATLAB 2021b with Gurobi 9.1.2 is utilized to solve the MILP problems, and the gap for the MILP problems is set to 0.0001. The specific network configurations corresponding to the three different network cases considered in the calculations are shown in Figs. 6-8. The computation time for network case c is 5.34 seconds.

*B. Discussion of Network Slicing Cases*

A comparison of the three network cases illustrates the advantages of the network slicing scheme proposed in this paper.

In Figs. 6-8, the devices and links outlined in red and gray represent that they are failure and communication interruption states, respectively. As shown in Fig. 6, when the traditional PDS communication networks are faced with the above problem scenarios, the following flexibility and resilience problems arise. First, with the single tree topology, if switch 3 (network switch with red outline in Fig. 6) fails, all subordinate facilities lose communication with the operation center. Consequently, the operation center loses monitoring and control capabilities over the electricity facilities managed by these terminal devices (terminal devices marked in gray). What's more, the system lacks automatic recovery capabilities, requiring manual intervention which leads to prolonged recovery time. Additionally, due to traditional private networking models, even though device 17 is proximate to switch 9, it must connect to network A (the network in the blue zone) via a long-distance link (the green link between device 17 and switch 6), leading to significant communication link construction costs. Similarly, the newly added terminal device 61 near switch 11 requires a long-distance link (the green link between device 61 and switch 8) for accessing the service B network, resulting in large construction expenses.

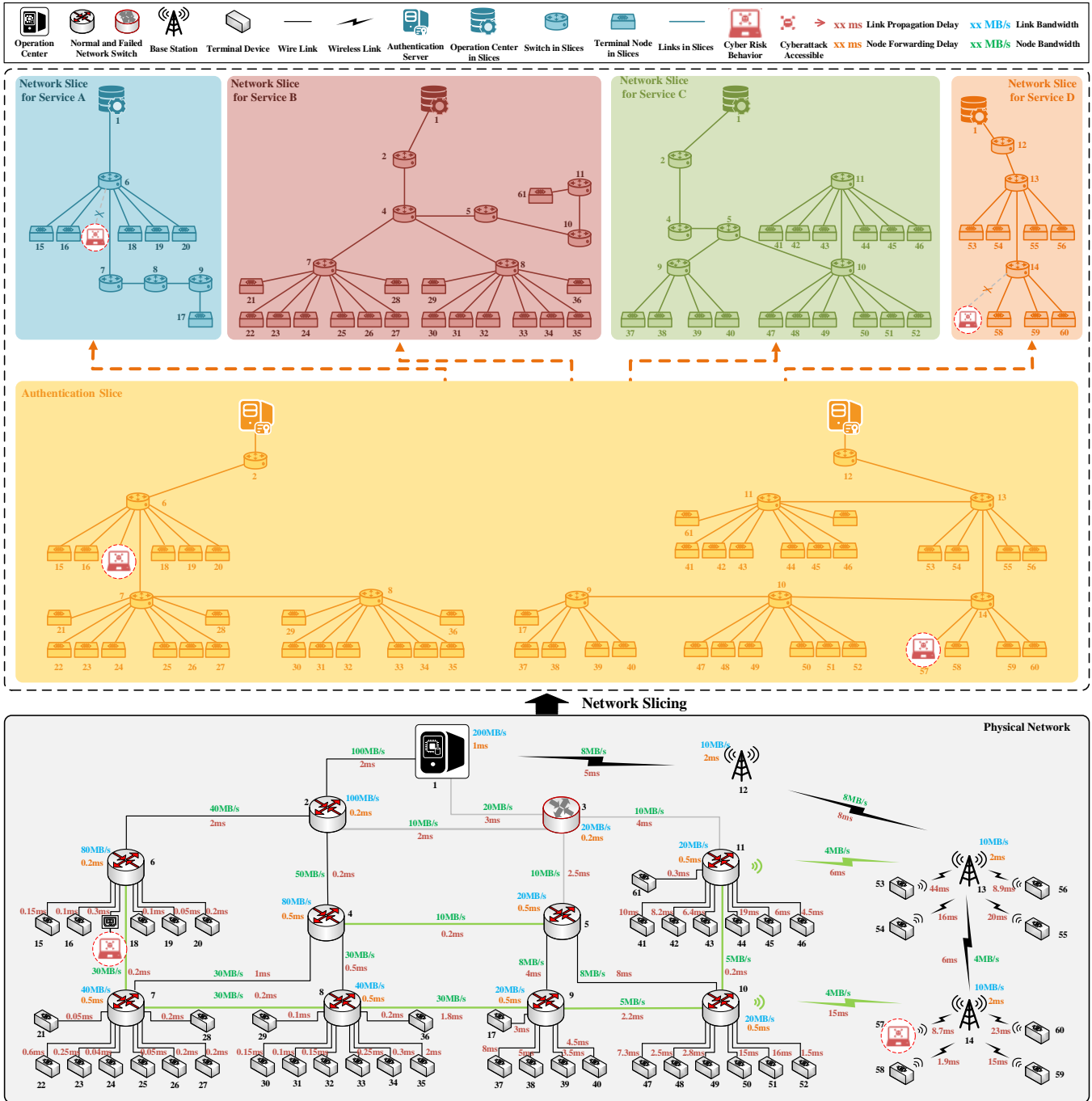

**Fig. 8.** The integrated network sliced by means of the proposed authentication slice-based dynamic network slicing method in this scenario.

However, network slicing in the integrated network addresses the above issues of flexible resource allocation, as shown in Figs. 7 and 8. First, the network integration provides deployable network resources for restoring communication links in the face of network failures. Network slicing of the integrated network allocates network resources in accordance with dynamic service demands. The standby links between switches 4 and 5, 8 and 9, and 10 and 14 (the green links between forwarding nodes in Figs. 7 and 8) can be used to restore communication for terminal devices and the operation center that are affected by switch 3. As illustrated in Figs. 7 and 8, utilize the remaining bandwidth of the link between base station 12 and the operation center, and the link between switch 2 and the operation center, we establish network slices for service C (shown as the green areas in Figs. 7 and 8) to restore communication for service C's terminal devices. This approach leads to effective fault recovery and improved resource utilization. Additionally, the newly added node 61 can now be directly connected to the adjacent switch 11. Network slicing for service B is established by allocating the remaining bandwidth of links between switch 11 and other forwarding nodes, enabling device 61 to communicate with the operation center through the redivided slice and thus saving on link construction costs. Similarly, the displaced node 17 is connected to the adjacent switch 9, with network slicing executed to facilitate a connection between node 17 and the operation center, further reducing link construction costs. These results illustrate that the network integration and slicing method proposed in this paper promotes network resource sharing, bolsters network resilience, and reduces network expansion costs.

Nevertheless, cybersecurity issues remain prominent. As depicted in Fig. 6, when device 17 is relocated and its network interface is not manually deactivated promptly, the interface retains network access and can serve as a port for unauthorized access. Moreover, illegitimate devices can easily masquerade as terminal devices (e.g., node 57 in this scenario), posing significant cybersecurity risks. As depicted in Fig. 6, illegitimate devices accessing through the original port of terminal 17 and those using terminal 57's port, disguised as terminal 57, can readily access networks A and D. This enables them to communicate with terminal devices and servers of the operation center across these networks, and to acquire network data. Such unauthorized access could lead to illegitimate activities, including attacks on operation center servers and the malicious control of critical infrastructure (e.g., generators and automatic breakers in key power lines), potentially resulting in significant power outages.

The cybersecurity of the PDS will not be greatly improved by the direct application of network slicing techniques such as those represented in Fig. 7. Without stringent device access mechanisms, networks utilizing conventional slicing methods remain vulnerable to cyberattacks. Illegitimate network access targeting service A and service D, as shown in Fig. 7, can similarly compromise the server and all terminal devices (terminal devices 15-20, and 53-60) and network facilities (network switches 6-9 and base stations 12-14) associated with these services. Such attacks can readily penetrate service networks, thereby gaining communication capabilities and the ability to manipulate related facilities. In networks sliced using the CAC mechanism as shown in Fig. 7, the network integration actually enlarges the potential surface for cyberattacks. This interconnectedness, exemplified by links 7-8, 8-9, and 11-13, allows cyberattacks to compromise and disrupt multiple services, such as services B and C, through a single interface. In contrast, the automated device checking and identification mechanism of the authentication slice mitigates these risks. As demonstrated in Fig. 8, our authentication slice-based method effectively deactivates the service slice access of the port for the removed device 17 and isolates unauthenticated devices (disguised as device 57) within the authentication slice. In the latter scenario, due to the challenging cryptographic authentication mechanisms, illegitimate devices at ports 17 and 57, are isolated within the authentication slice. This isolation, inherent to slicing, prevents these devices from accessing the networks of services A-D. Consequently, it has become challenging for cyberattacks to illicitly obtain data or maliciously target the facilities of these services. This dynamic verification and risk isolation capabilities of the authentication slice significantly enhance the security of PDS communication networks. Additionally, this mechanism of automatic device authentication and slice adjustments facilitates rapid device access and management. Consequently, this method allows for secure plug-and-play operations of terminal devices and automatic fault recovery in networks in a secure manner.

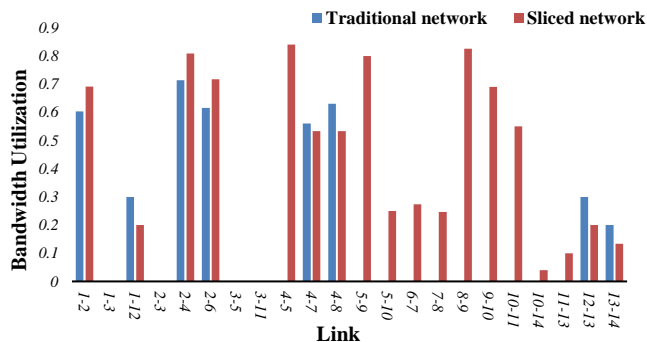

**Fig. 9.** Link bandwidth utilization in the traditional and sliced networks.

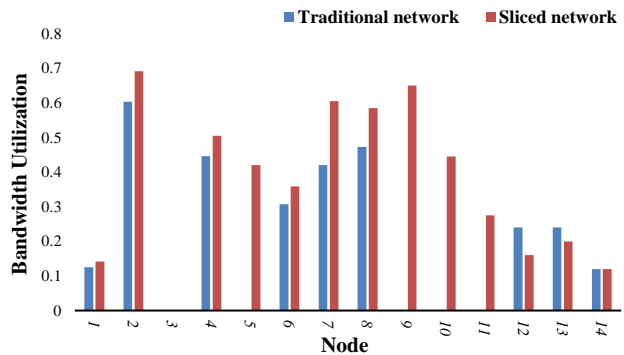

**Fig. 10.** Node bandwidth utilization in the traditional and sliced networks.

The results of bandwidth utilization for links and nodes in both traditional and sliced networks are depicted in Figs. 9 and 10. In these figures, the blue bars indicate the bandwidth utilization for the traditional networks, while the red bars represent that of the integrated network using our proposed network slicing method. Utilizing the remaining bandwidth in the sliced network to restore interrupted links, we observed an increase in bandwidth utilization for approximately 59.1% of the links and 71.4% of the nodes. However, some network links (links 2-4, 4-5, 5-9, and 8-9) and nodes (nodes 2, 7, and 9) approach saturation and require optimization.

*C. Network Robustness Evaluation and Optimization*

Based on the above network architectures, we calculate the numbers of terminal devices at which service is affected in cases of failure at each forwarding node and at each link between forwarding nodes, and the results are plotted in Figs. 11 and 12, respectively.

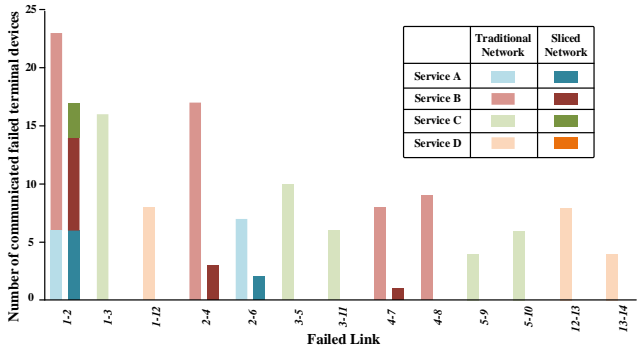

**Fig. 11.** Service impacts of link failures in the traditional and sliced networks.

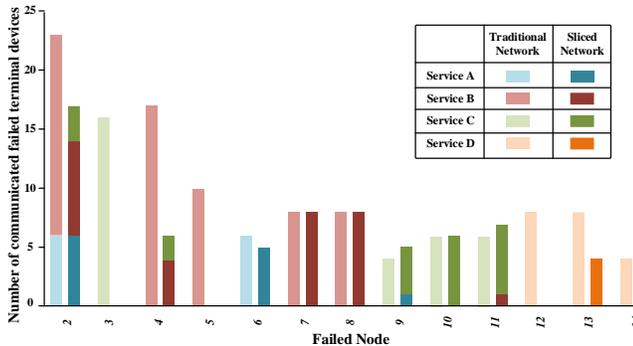

**Fig. 12.** Service impacts of node failures in the traditional and sliced networks.

In Figs. 11 and 12, light and dark bars in four different colors depict the number of terminal devices of four services that experience communication interruptions with the operation center due to node and link failures in the traditional and sliced networks, respectively, across the four services.

The assumed FRA parameters of the links and nodes are listed in Table III. Additionally, the FRA parameters of the links added to the structure for network integration are shown in Table IV.

TABLE III
FRA PARAMETERS OF NODES AND LINKS

| Node | FRA parameter | Link | FRA parameter |
|---|---|---|---|
| 2 | 12 | 1–2 | 12 |
| 3 | 8 | 1–3 | 4 |
| 4 | 4 | 1–12 | 4 |
| 5 | 4 | 2–4 | 8 |
| 6 | 4 | 2–6 | 4 |
| 7 | 4 | 3–5 | 4 |
| 8 | 4 | 3–11 | 4 |
| 9 | 2 | 4–7 | 4 |
| 10 | 2 | 4–8 | 4 |
| 11 | 2 | 5–9 | 4 |
| 12 | 2 | 5–10 | 4 |
| 13 | 2 | 12–13 | 2 |
| 14 | 2 | 13–14 | 2 |

TABLE IV
FRA PARAMETERS OF ADDED LINKS IN THE SLICED NETWORK

| Link | FRA parameter | Security level | Link | FRA parameter | Security level |
|---|---|---|---|---|---|
| 2–3 | 4 | 5 | 9-10 | 4 | 3 |
| 4–5 | 4 | 3 | 10-11 | 4 | 3 |
| 6–7 | 8 | 3 | 10-14 | 4 | 2 |
| 7–8 | 8 | 3 | 11-13 | 2 | 2 |
| 8–9 | 4 | 3 | / | / | / |

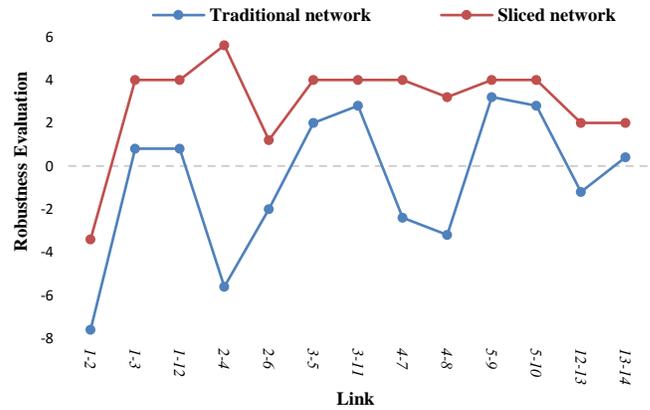

**Fig. 13.** Link robustness evaluation results for the traditional and sliced networks.

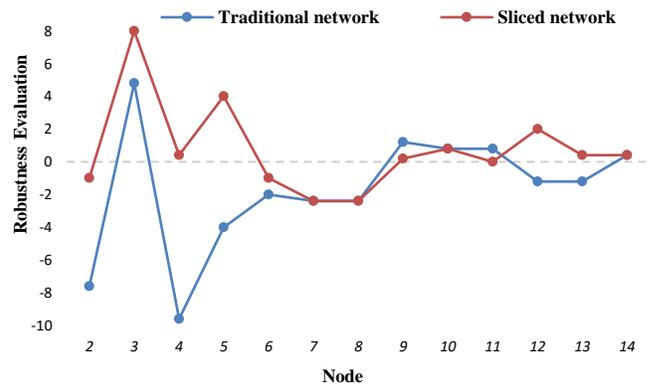

**Fig. 14.** Node robustness evaluation results for the traditional and sliced networks.

Based on the above parameters and the network structure, the robustness of the original and sliced networks for the above cases is evaluated using Algorithm 1. The results of the robustness evaluation for links and nodes are depicted in Figs. 13 and 14, respectively. In these figures, the red and blue curves represent the robustness assessment outcomes for links and nodes in the traditional and sliced networks, respectively.

A comparison in Figs. 11-12 reveals that in the sliced network, 100% of links and 53.8% of forwarding nodes experience significantly less service disruption upon failure, underscoring the high reliability of the sliced network. Consequently, as illustrated in Figs. 13-14, the sliced PDS communication network exhibits greater robustness in the event of node and link failures compared to traditional tree-based communication networks.

Nevertheless, to improve the robustness of the network and to balance the loads, the above slicing-based network structure needs some optimization of the links and nodes. For network performance reasons, we set the network link and node bandwidth utilization constraint parameters $\chi^L$ and $\chi^N$ to 70% to prevent degradation in network data transmission performance and set the robustness constraint parameters $\mu^L$ and $\mu^N$ to 0 to improve the network's fault tolerance.

Tables V and VI show the optimization options for wired forwarding nodes (such as network switches) and wireless forwarding nodes (such as base stations), respectively, under reasonable assumptions, along with their cost, bandwidth (Mbps), and FRA parameters. The options for optimizing and adding links, along with their associated parameters are also shown in Tables VII and VIII, respectively.

TABLE V
OPTIMIZATION OPTIONS FOR WIRED COMMUNICATION NODES

| Option | Added bandwidth | Added FRA | Cost |
|---|---|---|---|
| 1 | 5 | 2 | 1 |
| 2 | 10 | 3 | 1.5 |
| 3 | 20 | 4 | 2 |
| 4 | 40 | 6 | 3 |
| 5 | 60 | 8 | 4 |
| 6 | 80 | 9 | 5.5 |
| 7 | 100 | 10 | 6 |

TABLE VI
OPTIMIZATION OPTIONS FOR WIRELESS COMMUNICATION NODES

| Option | Added bandwidth | Added FRA | Cost |
|---|---|---|---|
| 1 | 2 | 1 | 1.2 |
| 2 | 5 | 2 | 1.8 |
| 3 | 10 | 3 | 2 |
| 4 | 20 | 4 | 2.4 |

TABLE VII
OPTIMIZATION OPTIONS FOR COMMUNICATION LINKS

| Link | Added bandwidth | Added FRA | Cost |
|---|---|---|---|
| 1–2 | 20/60/100 | 4 | 4.2/6.2/8.2 |
| 1–3 | 5/10/20 | 3.3 | 2.5/3/3.5 |
| 1–12 | 2/4/8 | 2.8 | 1.38/1.85/2.55 |
| 2–3 | 5/10/20 | 3.3 | 2.39/2.89/3.39 |
| 2–4 | 20/25/50 | 3.2 | 2.82/3.22/4.82 |
| 2–6 | 10/20/40 | 3.1 | 2.11/2.71/3.41 |
| 3–5 | 5/10/20 | 3.2 | 1.87/2.47/3.07 |
| 3–11 | 5/10/20 | 3.2 | 1.9/2.5/3.1 |
| 4–5 | 5/10/20 | 3.3 | 2.45/2.95/3.45 |
| 4–7 | 10/20/40 | 3.1 | 2.34/2.94/3.74 |
| 4–8 | 10/20/40 | 3.1 | 2.1/2.7/3.5 |
| 5–9 | 2/4/8 | 1.9 | 1.28/1.68/2.08 |
| 5–10 | 2/4/8 | 1.9 | 1.25/1.65/2.05 |
| 6–7 | 10/20/40 | 3.1 | 2.1/2.7/3.5 |
| 7–8 | 10/20/40 | 3.1 | 1.95/2.55/3.35 |
| 8–9 | 10/20/40 | 3.1 | 2.05/2.65/3.45 |
| 9–10 | 5/10/20 | 3.0 | 2.5/3/3.5 |
| 10–11 | 5/10/20 | 3.0 | 2.27/2.75/3.25 |
| 10–14 | 2/4/8 | 1.8 | 1.8/1.95/2.1 |
| 11–13 | 2/4/8 | 1.8 | 1.8/1.95/2.1 |
| 12–13 | 2/4/8 | 1.8 | 1.45/1.65/1.85 |
| 13–14 | 2/4/8 | 1.8 | 1.7/1.85/2.0 |

TABLE VIII
OPTIONS FOR COMMUNICATION LINK ADDITION

| Link | Added bandwidth | Security level | Added FRA | Cost |
|---|---|---|---|---|
| 1–6 | 20/40/60 | 4 | 6 | 4.35/5.15/5.95 |
| 4–6 | 20/40/60 | 3 | 4 | 3.9/4.7/5.5 |
| 5–11 | 10/20/40 | 3 | 4 | 2.6/3.2/3.8 |
| 2–5 | 20/40/60 | 3 | 4 | 4.25/5.05/5.85 |
| 3–4 | 20/40/60 | 3 | 4 | 4.1/4.9/5.7 |
| 11–12 | 5/10/20 | 3 | 2 | 3.2/4/4.8 |

When these network link and node optimization options are introduced into the above robustness and load balancing optimization model, the lowest-cost network optimization scheme satisfying the network robustness and load balancing requirements that is obtained by solving the model is shown in Table IX below.

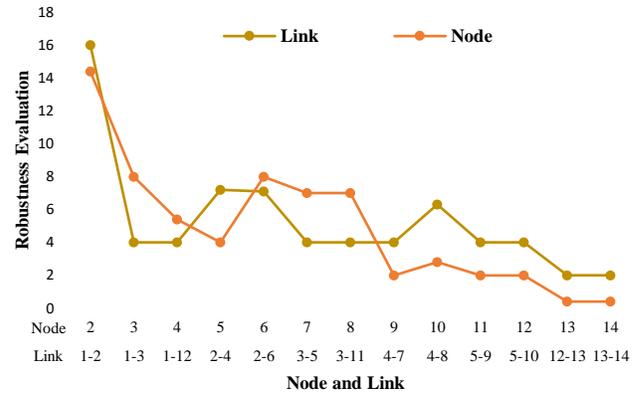

**Fig. 15.** Node and link robustness evaluation results for the optimized network.

TABLE IX
NETWORK OPTIMIZATION SOLUTION

| Operation | Component | Added bandwidth | Added FRA | Cost |
|---|---|---|---|---|
| Link optimization | 1–2 | 20 | 4 | 4.35 |
|  | 2–6 | 10 | 3.1 | 2.11 |
|  | 4–8 | 10 | 3.1 | 2.1 |
| Link addition | 1-6 | 60 | 6 | 5.95 |
| Forwarding node optimization | 2 | 20 | 4 | 2 |
|  | 4 | 10 | 3 | 1.5 |
|  | 6 | 60 | 4 | 4 |
|  | 7 | 10 | 3 | 1.5 |
|  | 8 | 10 | 3 | 1.5 |
|  | 10 | 5 | 2 | 1 |
|  | 11 | 5 | 2 | 1 |

Following the update of the integrated network based on the aforementioned optimization solution, Fig. 15 presents the recalculated robustness results for nodes and links in the optimized PDS network. In this figure, the yellow and orange curves represent the robustness assessment results for link and node failures in the sliced network, respectively. These results indicate that the robustness assessments for all nodes and links, as computed by the model, exceed the established robustness criteria.

Comparing Fig. 15 with Figs. 13 and 14 shows that the robustness of the optimized network is significantly better than that of the preoptimized network. This indicates that the optimized network has fewer services affected in case of the same failure of nodes and links. Thus, the proposed optimization solution improves the resilience of the network in the event of network component failures. Moreover, the presented optimization model enables the selection of the most economical solution for network load balancing and robustness optimization among the available optimization options when the network structure is afflicted by bandwidth saturation or lack of robustness. In this way, communication networks can be optimized in accordance with changes in communication service requirements that occur during PDS development.

## VI. CONCLUSION

This paper presents a novel multiservice network integration and network slicing scheme for PDS

communication networks. Considering the shortcomings in the flexibility of the traditional PDS communication based on multiple separate networks, network integration and slicing are first performed to enable network resource sharing and network resilience in PDSs. Then, to enhance network security and device interfacing capabilities, an authentication slice-based network slicing scheme with a device interfacing mechanism is also proposed. The proposed authentication slice-based authentication and slicing mechanism helps achieve isolation of unauthorized devices, plug-and-play support for terminal devices and dynamic adjustment of network slices. Moreover, to improve network performance in response to PDS development and evolving PDS services, evaluation and optimization models for the load balancing and robustness enhancement of PDS communication networks are proposed. Ultimately, the proposed methods are evaluated and demonstrated to excel in resource utilization, fault recovery, terminal device plug-and-play support, load balancing, and network robustness. Future initiatives will concentrate on strengthening authentication server defenses against DDoS cyberattacks and improving network slicing techniques for proactive resource partitioning to accommodate the evolving demands of various PDS services.